\documentclass[aps,prd,twocolumn,nofootinbib]{revtex4}
\usepackage{graphicx,bm}
\usepackage{amsmath,amssymb,mathrsfs}

\begin{document}
\title{Charmonium sum rules applied to a holographic model}
\author{Paul~M.~Hohler}
\email{pmhohler@uic.edu} \affiliation{Department of Physics,
University of Illinois, Chicago,  Illinois 60607-7059, USA} 


\begin{abstract}
The heavy-quark QCD sum rules are applied to a model of charmonium
based upon the gauge/gravity duality. We find that there is strong
agreement between the moments of the polarization function
calculated from the holographic model and the experimental data
suggesting that the model is consistent with the heavy-quark QCD
sum rules at zero temperature.
\end{abstract}

\maketitle

\section{Introduction}
A new model of charmonium in the context of the gauge/gravity
duality was constructed at zero temperature using spectral data so
that the model exactly reproduced the masses and decay constants
of the ground state, $J/\psi$, and the first excited state,
$\psi'$ \cite{Grigoryan:2010pj}. This model was then considered at
finite temperature, where the dissociation temperature of
charmonium was investigated. In this paper, we will examine the
zero temperature model in the context of heavy-quark QCD sum rules
\cite{Shifman:1978bx,Shifman:1978by,Novikov:1977dq}. As a
reference point, the results will be compared with a similar
analysis of a different holographic model of charmonium
\cite{Fujita:2009wc,Fujita:2009ca}.

QCD sum rules were first developed by Shifman, Vainshtein, and
Zakharov (SVZ) \cite{Shifman:1978bx,Shifman:1978by}. They provide
a systematic way to compare theory and experiment and have been
used in a wide variety of systems. QCD sum rules are based upon
the two-point vector-vector correlation function of the
heavy-quark (charm) current; at zero temperature this correlation
function can be expressed in terms of one function, the
polarization function, $\Pi^{(c)}(Q^2)$,
\begin{equation}
\int \! d^4 \! x \,e^{i q x}\, \langle \!J_\mu(x) J_\nu(0)\rangle
= \left(q_\mu q_\nu -q^2 g_{\mu\nu}\right) \Pi^{(c)}(Q^2),
\end{equation}
where the charm current is defined as $J^{\mu} =
\bar{c}\gamma^{\mu}c$ and $Q^2 = -q^2$. The charm polarization
function can further be expressed in terms of a dispersion
relation
\begin{equation} \label{eq:disp}
-\frac{d}{dQ^2} \Pi^{(c)}(Q^2) = \frac{1}{12 \pi^2 Q_c^2} \int \!
\frac{R_c (s) ds}{(s+Q^2)^2},
\end{equation}
where $Q_c=2/3$ is the charm quark electric charge, and $R_c(s)$
is the imaginary part of the polarization function and is related
to the cross section $\sigma_c$ as
\begin{equation}
R_c = \frac{3 s \sigma_c}{4 \pi \alpha^2}.
\end{equation}

QCD sum rules states that the cross section, and thereby the
polarization, can be formulated either from experimental data or
theoretically from an operator product expansion (OPE). Since both
of these should describe the same physics, the polarization
function one obtains from either method should be the same. This
leads to the construction of equations between the polarization
function calculated using either the OPEs or the experimental
data. These equations can be solved for various parameters of the
OPEs.

As pointed out by SVZ, the OPE side of the sum rule becomes
tractable when one considers a regime where asymptotic freedom
allows one to use perturbative QCD. In general, this is valid for
energies which satisfy $Q^2 + 4 m_q^2 \gg \Lambda_{\rm QCD}^2$.
For light quarks, this is only satisfied when $Q^2 \gg
\Lambda_{\rm QCD}$. However, in the case of heavy quarks, such as
charm, the relation holds even when $Q^2=0$. SVZ used this fact to
construct QCD sum rules which are unique to heavy-quark
systems\footnote{Though in this note, we focus on the heavy quark
sum rules at $Q^2=0$, heavy quark sum rules have been used at
$Q^2>0$ based upon the initial work of
Ref.~\cite{Reinders:1984sr}}. In the region of small $Q^2$, it is
useful for comparisons to introduce the moments of the
polarization function defined as
\begin{equation} \label{eq:moment}
\begin{split}
\mathscr{M}_n &\equiv \frac{1}{12 \pi^2 Q_c^2} \int \!
\frac{R_c(s) ds}{s^{n+1}} \\&\quad\quad=
\left.\frac{1}{n!}\left(-\frac{d}{dQ^2}\right)^n
\Pi^{(c)}(Q^2)\right|_{Q^2=0}.
\end{split}
\end{equation}
One can then relate each moment between the OPE side and the
experimental data rather than the entire polarization function. In
the original work by SVZ \cite{Shifman:1978bx,Shifman:1978by},
using this method of comparison, they were able to calculate the
heavy-quark mass and show the need for nonperturbative corrections
to have better agreement with the data.

Recently, the development of gauge/gravity dualities based upon
the AdS/CFT correspondence
\cite{Maldacena:1997re,Gubser:1998bc,Witten:1998qj} have led to
holographic models which provide a new method of calculating the
polarization function. Therefore it is natural to apply QCD sum
rules as a way to understand, constrain, and evaluate possible
holographic models. The OPE expansion has already been used for
light quark holographic systems in order to show that the AdS
metric is sufficient to reproduce the leading large $Q^2$ behavior
\cite{Erlich:2005qh,Da Rold:2005zs}, and agreement with the OPE
expansion has been a guiding principle in the construction of many
light quark holographic models among which includes
\cite{Hirn:2005vk,Csaki:2006ji,Afonin:2009pd,Cappiello:2009cj,Hambye:2006av}.
Yet this is the first time that these techniques will be applied
to a holographic model with heavy quarks.

In the next section, the gauge/gravity dual models of charmonium
\cite{Grigoryan:2010pj,Fujita:2009wc} are introduced. In
Sec.~\ref{sec:moments}, the moments of the polarization function
will be calculated from perturbative QCD as well as the
holographic model \cite{Grigoryan:2010pj}. This will be followed
in Sec.~\ref{sec:results} by the results of the comparison of the
moments between QCD and the holographic models, where the
holographic models will take the role of the OPE side in the sum
rules. Finally, we will conclude, in Sec.~\ref{sec:disc}, with a
discussion of the results.

\section{Charmonium model from gauge/gravity duality}
\label{sec:holo}

Key features relevant for the current discussions of holographic
models of charmonium will be presented here. A more detailed
exposition of these models can be found in
Refs.~\cite{Grigoryan:2010pj} and \cite{Fujita:2009wc}. The model
of Ref.~\cite{Grigoryan:2010pj} will be referred to as the ``shift
and dip" model while the model of Ref.~\cite{Fujita:2009wc} will
be referred to as the ``rescaled $\rho$" model. The basic
construction of the two models is similar, but the latter does not
reproduce as many phenomenologically relevant features as the
shift and dip model. The differences between the two models will
be pointed out, and the two models will each be compared with the
heavy-quark QCD sum rules in the subsequent sections.

In the spirit of the holographic approach, it is assumed that the
generating functional of the heavy-quark vector current $J^{\mu}$
can be represented by the effective action obtained by integrating
over a bulk 5D gauge field $V_M$ (dual to the current) at a given
fixed boundary value (equal to the source of the current). The
action for the 5D gauge field is given by
\begin{equation}
  \label{eq:action-5d}
  S = -\frac{1}{4g_5^2}\int d^5x \sqrt{g}\, e^{-\Phi}V_{MN}V^{MN},
\end{equation}
where $g_5^2$ is the 5D gauge coupling and $V_{MN} = \partial_M
V_N - \partial_N V_M $. The two-point current correlator is given
by the linear response of the field $V_M$ to an infinitesimal
perturbation of its boundary condition.

The conformally flat representation for the 5D background metric
$g_{MN}$ with 4D Lorentz isometry is chosen:
\begin{equation}
  \label{eq:ds2}
  ds^2\equiv g_{MN}\,dx^Mdx^N
=e^{2A(z)}\left[ \eta_{\mu\nu}dx^\mu dx^\nu - dz^2  \right] \ ,
\end{equation}
where $\eta_{\mu\nu}={\rm diag}(1,-1,-1,-1)$ is the Minkowski
metric tensor. The effect of confinement is represented by the
nontrivial background profile of the scalar field $\Phi$ in
Eq.~(\ref{eq:action-5d}) in the same way as it is done in the
soft-wall model with dilaton background in
Ref.~\cite{Karch:2006pv}.

Following the rules of the holographic correspondence, the
generating functional for correlation functions of the heavy-quark
current  can be calculated by evaluating the action at its
extremum for given boundary conditions. The extremum is given by
the solution of the equations of motion, which in $V_5=0$ gauge
read
\begin{equation} \label{eq:eom-V}
\partial_z[e^{B\left(z\right)} \partial_z V] + q^2 e^{B\left(z\right)}V = 0 \ ,
\end{equation}
where $V$ is any of the three components $V_\perp$ of $V_\mu(z,q)$
transverse to 4-vector $q^\mu$ ($q^\perp=0$) and
\begin{equation}
  \label{eq:B-Phi-A}
   B=A-\Phi\,.
\end{equation}

Discrete values of $q^2=m_k^2$, for which Eq.~(\ref{eq:eom-V})
possesses a normalizable solution $V=v_k(z)$ satisfying the
boundary condition $V|_{z=0}=0$, correspond to the masses $m_k$ of
the charmonium states, $k=1,2,\ldots = J/\psi,\psi',\ldots\ $. We
normalize such solutions as
\begin{equation}
\label{eq:norm}
  \int_0^\infty dz\, e^{B(z)}v_k(z)^2=1\ .
\end{equation}

The current-current correlator can be calculated according to the
well-known prescription of Ref.~\cite{Son:2002sd,Herzog:2002pc},
by
\begin{equation}
  \label{eq:G_R-V}
  G_R(q) = -\frac1{g_5^{2}}\,  e^B{V'(z,q)}\bigg|_{z=\epsilon}=
  -\frac1{g_5^{2}}\,\frac{V'(\epsilon,q)}{\epsilon},
\end{equation}
where $\epsilon\to0$ is an ultraviolet regulator and $V(z,q)$
is the non-normalizable solution of Eq.~(\ref{eq:eom-V}) with
boundary conditions:
\begin{equation}
  \label{eq:bc-V}
  \begin{split}
&    V(\epsilon,q)=1\ ;\quad
\\
&    V(z,q)\xrightarrow{z\to \infty}
    0\ .
  \end{split}
\end{equation}
The polarization function can then be determined by
\begin{equation} \label{eq:polar}
\Pi^{(c)}(Q^2) = \left.-\frac{1}{g_5^2Q^2}e^{B}
V'(z,q)\right|_{z=\epsilon},
\end{equation}
where $Q^2 = -q^2$. The ultraviolet behavior of the
current-current correlator is required to be conformal, {\it viz.}
$G_R(q)\sim q^2\log(-q^2)$ as $q^2\to-\infty$, which translates
into
\begin{equation}
  \label{eq:B-small-z}
  e^{B(z)}\xrightarrow{z\to0} z^{-1}\ ,
\end{equation}
and matches that of QCD, which
fixes~\cite{Son:2003et,Erlich:2005qh,Da Rold:2005zs}
\begin{equation}
  \label{eq:g_5-Nc}
 g_5^2=12\pi^2/N_c\ .
\end{equation}

By performing a Liouville transformation,
\begin{equation}
  \label{eq:Liouville-0}
  \Psi = e^{B(z)/2} V\ ,
\end{equation}
we can bring Eq.~(\ref{eq:eom-V}) to the canonical
Schr\"odinger-like form,
\begin{equation}
  \label{eq:Schroedinger-0}
 - d^2\Psi/dz^2 + U(z)\Psi = q^2\Psi\ ,
\end{equation}
with the holographic potential given by
\begin{equation}
  \label{eq:U-B-0}
  U(z) =
\frac{B''(z)}2 + \left(\frac{B'(z)}{2}\right)^2\ .
\end{equation}

At this point the two holographic models differ. The rescaled
$\rho$ model chooses the metric warp factor to be $A(z) =
-\log(z)$ and the dilaton profile to be $\Phi(z) = a^2 z^2$. This
is precisely the form of the standard soft-wall model
\cite{Karch:2006pv}. This leads to a model of one parameter, {\it
viz.} $a$, which is determined by fixing the mass of the $J/\psi$.
Therefore the rescaled $\rho$ model correctly reproduces the
$J/\psi$ mass (by construction), but the $J/\psi$ decay constant
and the $\psi'$ mass and decay constant are all off by nearly
$20\%$.

On the other hand, in the spirit of the bottom-up approach, the
shift and dip model chooses the function $B$ so as to satisfy the
spectroscopic data associated with $J/\psi$ and $\psi'$. It is
assumed that such a background arises dynamically, but no attempt
to model the corresponding dynamics is made. To this end, a
holographic potential $U(z)$ is chosen:
\begin{equation}
    \label{eq:piecewise-U}
    \begin{split}
    U(z) &= \frac3{4z^2}\,\theta(z_d-z) \\&\quad+ \left((a^2 z)^2 + c^2\right)
    \,\theta(z-z_d) - \alpha\delta(z-z_d).
    \end{split}
  \end{equation}
There are four parameters in this potential, which will be used to
fit four experimental data points: the masses and the decay
constants of $J/\psi$ and $\psi'$. This potential may seem rather
unusual, but it was chosen to reproduce the correct expected
behavior at both small and large $z$ as well as adding features,
{\it viz.} the $c^2$ ``shift" term and the delta function ``dip"
term, which facilitated the matching of the parameters to the
physical constraints. The matched parameters were found to be
\begin{equation}\label{eq:parameters}
  \begin{split}
    &a=0.970\mbox{ GeV},\  c=2.781\mbox{ GeV},\\
    &\alpha=1.876\mbox{ GeV},\ z_d^{-1}=2.211\mbox{ GeV} .
  \end{split}
\end{equation}

The potential in Eq.~(\ref{eq:piecewise-U}) can be, certainly,
improved by applying further constraints. One goal of this paper
is to determine how well this potential holds up to the additional
constraints imposed by the heavy-quark QCD sum rules.

Having established $U(z)$, the polarization function can be found
by first solving Eq.~(\ref{eq:Schroedinger-0}) for the function
$\Psi(z)$. Equation (\ref{eq:U-B-0}) can be solved for the
function $B(z)$. Therefore the vector field can be calculated from
$B(z)$ and $\Psi(z)$ using Eq.~(\ref{eq:Liouville-0}). Finally the
polarization is calculated from Eq.~(\ref{eq:polar}) while its
moments can be generated from Eq.~(\ref{eq:moment}).

\section{Calculation of moments}
\label{sec:moments}

In  this section, the moments will be calculated directly from the
polarization function for the QCD OPE expansion and the shift and
dip holographic model. The calculation of the polarization
function using the rescaled $\rho$ model is presented in Appendix
\ref{sec:rho}. The method to calculate the same moments from
spectroscopic data will also be explored. A more general
calculation of the moments for an arbitrary holographic model can
be found in Appendix \ref{sec:general}.

\subsection{OPE expansion} \label{sec:ope}
To calculate the moments for the QCD OPE expansion, we will
concentrate on the leading unit operator term. Nonperturbative
corrections, due primarily to the gluon condensate, can be
included, but we will later show that this is unnecessary for our
analysis here. This term can be calculated using perturbative QCD
because of the heavy charm quark. At the one-loop level, the
imaginary part of the polarization function is given by
\begin{equation} \label{eq:rcope}
R_c(s) = \frac{2}{3}v (3-v^2)\ \theta(s-4m_c^2),
\end{equation}
where the heavy-quark velocity is given by $v = (1-4
m_c^2/s)^{1/2}$, and $m_c$ is the charm quark mass. This
expression can be used in the dispersion relation of
Eq.~(\ref{eq:disp}) to calculate the polarization function, or it
can be used in Eq.~(\ref{eq:moment}) to find it moments. This
leads to the familiar result of \cite{Shifman:1978by} for the
moments,
\begin{equation}
\mathscr{M}_n^{(0)} = \frac{3}{4\pi^2} \,
\frac{2^n(n+1)(n-1)!}{(2n+3)!!} \,\frac{1}{(4m_c^2)^n},
\end{equation}
where superscript $(0)$ refers to the fact that this is the
leading order calculation. Note that charm quark mass is a free
parameter of the OPE.

The leading $\alpha_s$ correction to the moments can also be
calculated. With the inclusion of this correction the moments are
given by
\begin{equation}
\mathscr{M}_n = \mathscr{M}_n^{(0)}+\alpha_s \mathscr{M}_n^{(1)},
\end{equation}
where
\begin{equation} \label{eq:alphamoment}
\begin{split}
\mathscr{M}_n^{(1)} &=
\mathscr{M}_n^{(0)}\left(\frac{4\sqrt{\pi}}{3}\frac{\Gamma(n+3/2)}{\Gamma(n+1)}
\frac{1-(3n+3)^{-1}}{1-(2n+3)^{-1}} \right. \\
&\left.-\frac{1}{2}\pi+\frac{3}{4\pi}-\frac{4n \log 2}{\pi} \right. \\
&\left.-\frac{2}{3\sqrt{\pi}}\frac{2\pi^2-3}{4\pi}\frac{\Gamma(n+3/2)}{\Gamma(n+2)}\frac{1-2(3n+6)^{-1}}{1-(2n+3)^{-1}}\right).
\end{split}
\end{equation}

\subsection{Shift and dip model} \label{sec:dip}
In this section, the procedure discussed at the end of
Sec.~\ref{sec:holo} will be used to determine the polarization
function for the shift and dip model. The moments will not be
explicitly calculated, but can be found from Eq.~(\ref{eq:moment})
using the polarization function. The results will be expressed in
terms of the parameters $a$, $c$, $\alpha$, and $z_d$ whose
numerical values can be found in Eq.~(\ref{eq:parameters}). First,
one needs to calculate $B(z)$; this is done by solving
Eq.~(\ref{eq:U-B-0}) with the holographic potential giving by
Eq.~(\ref{eq:piecewise-U}). This results in
\begin{equation}
e^{B(z)/2} = \left\{\begin{array}{ll}
\frac{1}{\sqrt{z}}+c_1\,\frac{z^{3/2}}{z_d^2} & z<z_d \\
c_2\, 2^{\frac{a^2+c^2}{4a^2}} e^{-\frac{1}{2}a^2 z^2}
H_{\nu_1}\left(a z\right)&z>z_d
\end{array}\right. ,
\end{equation}
where $H_\nu(x)$ is the Hermite polynomial,
$\nu_1=-\frac{1}{2}-\frac{c^2}{2a^2}$, and $c_1$ and $c_2$ are
integration constants given by,
\begin{equation}
\begin{split}
c_1 &= \frac{\left(1+2\alpha z_d+2a^2 z_d^2 \right)H_{\nu_1}(a
z_d)-2a z_d H_{\nu_2}(a z_d)}{\left(3-2\alpha z_d-2 a^2
z_d^2\right)H_{\nu_1}(a z_d)+2a z_d
H_{\nu_2}(a z_d)}, \\
c_2 &= 2^{\frac{1}{2}\nu_1} e^{\frac{1}{2}a^2
z_d^2}\frac{1+c_1}{\sqrt{z_d}\,H_{\nu_1}\left(a z_d\right)},
\end{split}
\end{equation}
where $\nu_1$ is as before and $\nu_2=\nu_1+1$. In order to
determine the bulk vector field, $V(z,q)$, one needs to solve
Eq.~(\ref{eq:Schroedinger-0}) and use Eq.~(\ref{eq:Liouville-0}).
This results in
\begin{equation}
V(z,q)=\left\{\begin{array}{ll}\frac{c_3\!(Q^2) \,q z J_1(q z)-\frac{\pi}{2} q z \,Y_1(q z)}{1+c_1 \left(\frac{z}{z_d}\right)^2}& z<z_d\\
c_4(Q^2) \frac{H_{\nu_{1q}}(a z)}{H_{\nu_1}(a z)}& z>z_d
\end{array}\right. ,
\end{equation}
where $c_1$ and $c_2$ are as before and $c_3(Q^2)$ and $c_4(Q^2)$
are given by
\begin{widetext}
\begin{equation}
\begin{split}
 c_3(Q^2)&=  \frac{\pi}{2}\left(\frac{-2 z_d q Y_0(q z_d) H_{\nu_{1q}}(a z_d)+Y_1(q z_d)\left(\left(1+2\alpha z_d+2 a^2 z_d^2\right)H_{\nu_{1q}}(a z_d)-2 a z_d H_{\nu_{2q}}(a z_d)\right)}
 {-2 z_d q J_0(q z_d)H_{\nu_{1q}}(a z_d)+J_1(q z_d)\left(\left(1+2\alpha z_d+2 a^2 z_d^2\right)H_{\nu_{1q}}(a z_d)-2 a z_d H_{\nu_{2q}}(a z_d)\right)}\right)\\
 c_4(Q^2)&=\frac{H_{\nu_1}(a z_d)}{H_{\nu_{1q}}(a
 z_d)}\,\frac{c_3(Q^2) \,q z_d J_1(q
z_d)-\frac{\pi}{2} \,q z_d \,Y_1(q z_d)}{1+c_1},
\end{split}
\end{equation}
\end{widetext}
with $q^2=-Q^2$, and $\nu_{1q}=\nu_1+\frac{q^2}{2a^2}$, and
$\nu_{2q}=\nu_{1q}+1$. From this formula for the vector field, the
polarization function can be found using Eq.~(\ref{eq:polar}),
\begin{equation} \label{eq:shiftpol}
\Pi^{(c)}(Q^2) = \frac{-1}{g_5^2}\!\left(\!\gamma_{\rm E} \!+
\log\!\left(\frac{i Q
\epsilon}{2}\right)\!-c_3(Q^2)-\frac{2c_1}{Q^2 z_d^2}\right).
\end{equation}
One can show that, for small $Q^2$, there are terms in $c_3(Q^2)$
which cancel both the $c_1$ term and the $\log(Q)$ term. This
expression also has the standard $\log(\epsilon)$ divergence from
the UV cutoff. This can be removed by renormalization such that
$\Pi^{(c)}(0)=0$. The moments can be generated from this
expression using Eq.~(\ref{eq:moment}), however a simple analytic
expression is not possible and therefore will not be presented
here.

\subsection{Spectroscopic data} \label{sec:spec}
The moments of the polarization function can also be determined
directly from the spectroscopic data. If one assumes that the
cross section can be constructed as a series of delta functions as
is done in Refs.~\cite{Novikov:1977dq,Shifman:1978by}, one for
each charmonium state, $R_c$ can be written as
\begin{equation} \label{eq:rcdelta}
\begin{split}
R_c(s) &= \frac{9\pi}{\alpha^2} \sum_{k} \delta(s-m_k^2)
\frac{\Gamma_k^{ee}}{m_k}\,s\\& = \frac{16 \pi^2}{3} \sum_k
\delta(s-m_k^2) \frac{f_k^2}{m_k^2}\,s,
\end{split}
\end{equation}
where the sum is over all charmonium states with $k$ labelling
each state, $\Gamma^{ee}$ is the electronic width, and $f$ is the
decay constant. The moments can then be calculated by using this
formula in Eq.~(\ref{eq:moment}),
\begin{equation}
\mathscr{M}_n=\sum_{k} \frac{f_k^2}{m_k^{2n+2}}.
\end{equation}
Since the mass of the excited states tends to grow faster than its
decay constant, the higher excited charmonium states are
suppressed for the higher moments. When considering the
experimental spectrum, the spectrum exhibits a continuum of states
above the $D\bar{D}$ threshold. SVZ approximated this continuum by
choosing $R_c(s)$ to be
\begin{equation} \label{eq:cont}
R_c(s) = \frac{4}{3} \theta(s-16 {\rm GeV}^2),
\end{equation}
above the threshold. The contribution to the polarization function
and the moments can as before be found by using
Eq.~(\ref{eq:cont}) in Eqs.~(\ref{eq:disp}) and (\ref{eq:moment}),
respectively.

The necessary spectroscopic data, {\it viz.} the masses, $m_k$ and
decay constants, $f_k$, can either come from the physical
experimental results \cite{pdg}, or it can be calculated from the
spectrum of either holographic model. Unlike the experimental
data, the holographic models contain only an infinite tower of
zero width excited charmonium states with no continuum. Therefore,
for good agreement between the moments of the holographic models
and the experimental spectrum, one would hope that this tower of
states can approximate the continuum of QCD.

The equivalence between the two methods to calculate the moments,
{\it i.e.} from the polarization function or from the spectrum, is
due to equivalent ways of expressing the polarization function.
One tends to use the method which is easier to calculate with. In
this paper, discussions will highlight the method using the
spectroscopic data since it is more natural to interpret the
results in the language of this method, though both techniques
yield the same numerical results.

\section{Results}
\label{sec:results}

To assess the consistency of the holographic model of charmonium
with the heavy-quark QCD sum rules, we will examine the moments of
the polarization function qualitatively and quantitatively.  The
moments calculated from the holographic model will be compared to
those calculated from the experimental spectroscopic data and from
perturbative QCD. In order to provide a baseline for the results
of the shift and dip model, the results using the rescaled $\rho$
model will also be presented.

To begin, we will use the techniques of SVZ to analyze the moments
of the polarization function. The polarization function's moments
will be calculated for each holographic model using each model's
spectral data as detailed in Sec.~\ref{sec:spec}. For both
holographic models, a sufficient number of excited states were
considered\footnote{For the shift and dip model, 400 states were
used, while for the rescaled $\rho$ model, one can analytically
calculate for infinite number of states.}. The moments of the
polarization function are also calculated from the first two
experimentally known charmonium states and a contribution for the
continuum as done in \cite{Novikov:1977dq}.

The ratio of the moments of the polarization function between each
holographic model and the experimental data are constructed and
presented in Table \ref{tab:ratio} with superscript A referring to
the shift and dip model and superscript B referring to the
rescaled $\rho$ model. The results indicate that the shift and dip
model is consistent with QCD at worst at the $20\%$ level with
even better agreement for the higher moments. This is quite
reasonable considering that there should be $1/N_c$ corrections to
the holographic model.

\begin{table}[htb]
\begin{tabular}{c|c|c}
\hline Moment n&$\mathscr{M}_n^{ A}/\mathscr{M}_n^{\rm
Expt}$&$\mathscr{M}_n^{B}/\mathscr{M}_n^{\rm Expt}$\\ \hline
1&0.79&0.54\\
2&0.91&0.57\\
3&0.96&0.59\\
4&0.98&0.60\\
5&0.99&0.61\\
\hline
\end{tabular}
\caption{Ratio of the polarization function moments calculated
using either holographic model to the moment calculated using QCD.
The label A refers to the shift and dip model while the label B
refers to the rescaled $\rho$ model.} \label{tab:ratio}
\end{table}

The better agreement between the shift and dip model and QCD for
the higher moments can be understood by the fact that the higher
excited states are suppressed for these moments as discussed in
Ref.~\cite{Shifman:1978by,Novikov:1977dq}. Therefore the largest
contribution for both the holographic model and QCD comes from the
$J/\psi$ and $\psi'$ states which are identical in both cases by
construction. For example, the $J/\psi$ and $\psi'$ contribute
only $75\%$ to the first moment in the shift and dip model while
this climbs to $99\%$ for the fifth moment. Therefore the larger
discrepancy seen in the first moment is an indication that the
tower of excited states in the holographic model only marginally
approximates the QCD continuum.

The rescaled $\rho$ model also has improved agreement with QCD for
the higher moments because of the increase in the contribution of
the $J/\psi$ to these moments. However, the differences seen
between the shift and dip model and the rescaled $\rho$ model in
Table \ref{tab:ratio} can be attributed to fact that the rescaled
$\rho$ model does not accurately reproduce the $\psi'$ state.

\begin{figure}[htb]
  \centering
  \includegraphics[width=.45\textwidth]{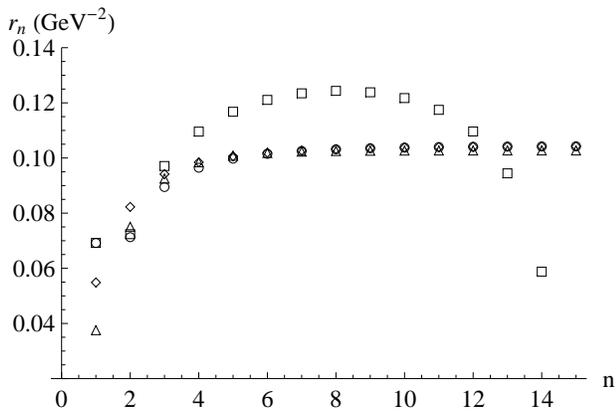}
  \caption{Ratio of consecutive moments calculated from
  perturbative QCD (squares), spectroscopic data (circles), the
  shift and dip model (diamonds), and the rescaled $\rho$ model
  (triangles).
  }
  \label{fig:ratio}
\end{figure}

Another useful way to compare the holographic model to the QCD OPE
and the experimental data is to look at the ratio of consecutive
moments, defined as
\begin{equation}
r_n = \frac{\mathscr{M}_{n}}{\mathscr{M}_{n-1}}.
\end{equation}
Figure \ref{fig:ratio} plots the results for $r_n$ from
perturbative QCD, the experimental data, and both holographic
models. The moments of the polarization function for the
experimental data and the holographic models are calculated as
described above. The polarization function's moments from
perturbative QCD are found from Eq.~(\ref{eq:alphamoment}) with
$m_Q =1.21$GeV. The charm quark mass was found by equating the
first moment calculated from the QCD OPE and the first moment
calculated from the experimental data and solving for the
heavy-quark mass. The points at $n=1$ correspond to the value of
the first moment multiplied by the factor $(4 \pi/3)^2$, as is
done by SVZ \cite{Shifman:1978by}. Again the discrepancy seen
between these values is indicative that the tower of excited
states of the shift and dip model does not reproduce the
experimental continuum well. One can see that for the highest
moments the holographic models agree significantly well with the
experimental data. Again this is not that surprising since the
highest moments have a larger contribution from the $J/\psi$
state. The discrepancy seen between the shift and dip model and
the rescaled $\rho$ model at the highest moments is because the
rescaled $\rho$ model only reproduces the $J/\psi$ mass and not
the decay constant. Furthermore, the perturbative QCD calculation
with only the leading $\alpha_s$ corrections, does not agree very
well with the experimental data for the higher moments. SVZ showed
\cite{Shifman:1978by} that this discrepancy can be resolved by
including the effects of the gluon condensate. The good agreement
between the holographic models would imply that they correctly
capture the nonperturbative physics of the charm system.

\begin{figure}[htb]
  \centering
  \includegraphics[width=.45\textwidth]{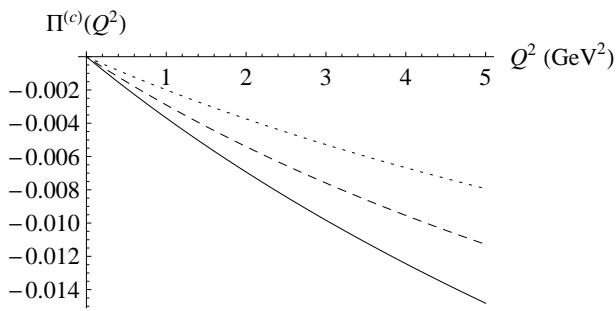}
  \caption{Polarization function $\Pi^{(c)}$ at small $Q^2$ calculated from the experimental data (solid line), the shift and
  dip model (dashed line), and the rescaled $\rho$ model (dotted
  line). All functions have been renormalized so $\Pi^{(c)}(0)=0$.
  }
  \label{fig:pi}
\end{figure}

The quantitative results from the moments can be summarized
graphically by plotting the polarization function directly. Figure
\ref{fig:pi} depicts the polarization function as a function of
$Q^2$ for small $Q^2$ calculated using the experimental
spectroscopic data (solid curve), the shift and dip model (dashed
curve), and the rescaled $\rho$ model (dotted curve). Each
function has been renormalized so that $\Pi^{(c)}(Q^2=0)=0$. The
curve from the experimental data is calculated from
Eq.~(\ref{eq:disp}) using Eqs.~(\ref{eq:rcdelta}) and
(\ref{eq:cont}). The shift and dip curve is a plot of
Eq.~(\ref{eq:shiftpol}) with the parameters chosen from
Eq.~(\ref{eq:parameters}) while the rescaled $\rho$ curve plots
Eq.~(\ref{eq:rhopolar}). Note, though not shown, the polarization
function determined from the perturbtative QCD calculation would
lie directly on top of the experimental curve. From the figure,
one can clearly see that the charmonium model with a shift and a
dip has better agreement with experimental data as compared with
the rescaled $\rho$ model. The polarization function calculated
from the shift and the dip model is in modest agreement with the
QCD calculation (within 21\%) for $Q^2 < 1 {\rm GeV}^2$, but
remains within 26\% agreement to $Q^2=10 {\rm GeV}^2$. This
uncertainty is associated with the discrepancy of the first moment
seen in Table \ref{tab:ratio}.

\section{Discussion}
\label{sec:disc}

In this paper, we have considered a holographic model of
charmonium and analyzed it in the context of heavy-quark QCD sum
rules. We have illustrated how to calculate the moments of the
polarization function for this model. The moments have then been
used to compare the holographic models to the results obtained
from the QCD OPE expansion and the experimental spectroscopic
data.

The results have shown that the holographic model of charmonium
presented in Ref.~\cite{Grigoryan:2010pj} agrees very well, more
than $80\%$ for the first moment and reaching above $90\%$ for the
higher moments, with heavy-quark QCD sum rules. This was shown via
graphical means and by directly examining the moments of the
polarization function.

This has been the first time that heavy-quark QCD sum rules have
been applied to a holographic model. It is not obvious {\it a
priori} that for the heavy-quark sector the agreement between the
holographic model and QCD for $Q^2<0$ translates into agreement
between the holographic model and QCD for $Q^2>0$. Nevertheless,
within reasonable uncertainties, this is exactly what we have
demonstrated here. The results of the shift and dip model have
been compared to another holographic model as a baseline
comparison. This has illustrated that the added phenomenological
constraints imposed in constructing the shift and dip model
simultaneously improves the agreement with the QCD sum rules.
Clearly this agreement is enhanced by the fact that the moments
have a dominant contribution from the $J/\psi$ and $\psi'$ states.
However, the agreement, in particular for the first moment when
the contributions from $J/\psi$ and $\psi'$ to the moment are not
as significant, is nontrivial.

As one progresses to future holographic models of heavy-quark
systems, it will be important to understand how these systems
scale with the quark mass. In this way, one may be able to relate
aspects of the holographic models found in heavy-quark systems
back to their light mass cousins. Moreover, most of the
discrepancies between the shift and dip model and the QCD sum
rules seem to stem from discrepancies associated with the first
moment. Any future attempts to improve the agreement must address
this issue. Since each moment from the shift and dip model is
smaller than from the QCD OPE, to improve agreement one must
either increase the decay constants of the excited states or
decrease their masses. As discussed in the construction of the
shift and dip model in \cite{Grigoryan:2010pj}, these adjustments
may be accomplished by changing the holographic potential,
Eq.~(\ref{eq:piecewise-U}), by further increasing the region of
attraction, both in strength and width, in the area of $z \sim
m_q^{-1}$, or by possibly softening the soft wall, respectively.
Obviously, any future changes would still need to be consistent
with the spectroscopic data of the first two charmonium states.

In conclusion, we have demonstrated that the holographic shift and
dip model is consistent in a nontrivial manner with the QCD sum
rules at zero temperature.

\acknowledgments

P.M.H. would like to thank M.~Stephanov and H.~Grigoryan for
valuable suggestions essential to this work and discussions. The
work of P.M.H. is supported by the DOE Grant No.\
DE-FG0201ER41195.

\begin{appendix}
\section{Rescaled $\rho$ model} \label{sec:rho}
For the rescaled $\rho$ model, the moments can be calculated using
the procedure outlined at the end of Sec.~\ref{sec:holo}. However,
one does not need to solve any equations for the function $B(z)$
since it is provided from the soft-wall model. This function can
be expressed as
\begin{equation}
e^{B(z)} = \frac{1}{z}e^{-a^2 z^2},
\end{equation}
where $a$ is a parameter of the model which is determined by
requiring the model reproduce the mass of $J/\psi$ correctly.
Numerically it is $a=1.56$GeV. From this expression for $B(z)$,
one can determine $V(z)$ subject to the boundary conditions as
\begin{equation}
V(z,q) = \Gamma\!\left(1-\frac{q^2}{4 a^2}\right)
U\!\left(-\frac{q^2}{4 a^2},0,a^2 z^2\right),
\end{equation}
where $U(a,b,x)$ is the Tricomi confluent hypergeometric function.
This expression for $V(z,q)$ can then be used in
Eq.~(\ref{eq:polar}) to determine the polarization function for
this model. This results in
\begin{equation} \label{eq:rhopolar}
\Pi^{(c)}(Q^2) = -\frac{1}{2 g_5^2} \left[ \gamma_{\rm E}
+\psi\!\left(1+\frac{Q^2}{4 a^2}\right)\right],
\end{equation}
where $Q^2=-q^2$, $\gamma_{\rm E}$ is the Euler gamma, and
$\psi(x)$ is the polygamma (or digamma) function. This function
has already been renormalized so $\Pi^{(c)}(0)=0$. The moments of
the polarization function can be easily found using
Eq.~(\ref{eq:moment}) resulting in,
\begin{equation}
\mathscr{M}_n = \frac{1}{2 g_5^2} \frac{1}{(4a^2)^n}\, \zeta(n+1),
\end{equation}
where $\zeta(x)$ is the Zeta function.

\section{Determining the moments of the polarization function for a holographic
model}\label{sec:general}

In Sec.~\ref{sec:moments}, the moments of the polarization
function were calculated for a particular holographic model of
charmonium. In each case, one could solve the necessary
differential equation analytically, so an analytic expression for
the polarization function was possible. However, for an arbitrary
holographic model with a vector field this may not be the case. In
this Appendix, we will calculate the moments for this arbitrary
holographic model in terms of integral recursive relations. These
can be used when an analytic solution is possible, but will most
likely be more useful when one is required to rely on numerics.

To construct the moments, we will once again use the same
procedure that was outlined at the end of Sec.~\ref{sec:holo}.
However, since we wish to consider a large class of theories, we
will leave the function $B(z)$ unspecified. Therefore, we begin
with the equation of motion for the vector field of
Eq.~(\ref{eq:eom-V}),
\begin{equation} \label{eq:eom}
V^{\prime \prime}(z,Q^2) + B^\prime(z) V^\prime(z,Q^2) - Q^2
V(z,Q^2) = 0,
\end{equation}
where $q^2 = -Q^2$. As before, this equation is subject to the
boundary conditions of Eq.~(\ref{eq:bc-V}). Since we are
interested in the polarization function at small $Q^2$, we can
explore Eq.~(\ref{eq:eom}) in the same limit. Therefore we will
assume that $V(z,Q^2)$ can be expanded in a Taylor series of
$Q^2$, namely,
\begin{equation} \label{eq:series}
V(z,Q^2) = \sum_{n=0} \left(-Q^2\right)^n V_n(z).
\end{equation}
The functions $V_n(z)$ correspond to the coefficients in the
Taylor series. Note these should not be confused with the
eigenmodes of the differential equation. By inserting
Eq.~(\ref{eq:series}) into Eq.~(\ref{eq:eom}) and equating powers
of $Q^2$, one arrives at a difference equation for the functions
$V_n(z)$:
\begin{equation} \label{eq:eom2}
V_n^{\prime \prime} + B^\prime V_n^\prime = - V_{n-1},
\end{equation}
which can be solved by iterative means.

The 0th order expression,
\begin{equation}
V_0^{\prime \prime} + B' V_0^\prime = 0,
\end{equation}
can be easily solved,
\begin{equation}
V_0(z) = b_1 + b_2 \int_0^z e^{-B(z')} dz'.
\end{equation}
The integration constants, $b_1$ and $b_2$, can be determined from
the boundary conditions resulting in $V_0(z)=1$\footnote{Though
this expression does not explicitly satisfy the boundary condition
for $V(z,Q^2)$ at infinity, we have assumed that $V_n(z)$ should
at least be finite in this limit. The boundary condition for
$V(z,Q^2)$ would then be satisfied once the infinite sum in
Eq.~(\ref{eq:series}) is performed.}. For larger values of $n$,
the equation can be solved by a Green's function technique.
Therefore the solutions can be written as
\begin{equation} \label{eq:sol}
V_n(z) = \int_\epsilon^\infty dz^\prime G(z,z^\prime)
V_{n-1}(z^\prime),
\end{equation}
where $G(z,z')$ is defined to satisfy the differential equation,
\begin{equation} \label{eq:green}
\frac{\partial^2 G(z,z')}{\partial z^2} + B'(z) \frac{\partial
G(z,z')}{\partial z} = -\delta(z-z').
\end{equation}
To be consistent with the boundary conditions on $V(z,Q^2)$ and
the expression for $V_0(z)$, $G(z,z')$ must satisfy the boundary
conditions $G(\epsilon,z')=0$ and $G(z,z')$ is finite as $z
\rightarrow \infty$.

Equation (\ref{eq:green}) can be solved to the left and to the
right of delta function as
\begin{equation}
G(z,z') =\left\{\begin{array}{ll} b_3 + b_4 \int_\epsilon^z e^{-B(z'')} dz'' & z<z'\\
         b_5 + b_6 \int_{z'}^z e^{-B(z'')} dz'' & z>z' \end{array}
         \right. .
\end{equation}
The integration constants can be determined from the boundary
conditions and the matching conditions at the delta function
resulting in the Green's function to be expressed as,
\begin{equation} \label{eq:greensol}
G(z,z') =  e^{B(z')} \int_\epsilon^{z_<} e^{-B(z'')} dz'',
\end{equation}
where $z_<$ is the smaller of $z$ and $z'$. It is useful to note
the derivative of the Green's function with respect to $z$,
\begin{equation} \label{eq:gprime}
\frac{\partial G(z,z')}{\partial z} = \theta(z'-z)
\left(e^{B(z')-B(z)}\right),
\end{equation}
where $\theta$ is the unit step function.

Having determined the appropriate Green's function, $V_n(z)$ for
$n \geq 1$ can be generated from Eq.~(\ref{eq:sol}), in particular
$V_1(z)$ can be written as,
\begin{equation} \label{eq:v1}
\begin{split}
V_1 &= \int_\epsilon^\infty G(z,z')\, V_0(z')\, dz'\\
    &= \int_\epsilon^z \!
    \left(e^{B(z')}\int_\epsilon^{z'}e^{-B(z'')}dz''\right)dz'\\&\quad\quad+\int_z^\infty\!\left(e^{B(z')}\int_\epsilon^z
    e^{-B(z'')}dz''\right)dz'\\
    &= \int_\epsilon^z dz' e^{-B(z')} \int_{z'}^\infty
    e^{B(z'')}dz''.
\end{split}
\end{equation}
The notation here can be compactified some by defining the
functions $\alpha(z)$ and $\beta(z)$ as
\begin{equation}
\begin{split}
\alpha(z)&\equiv \int_\epsilon^z e^{-B(z')}\,dz',\\
\beta(z)&\equiv \int_z^\infty e^{B(z')}\,dz'.
\end{split}
\end{equation}
Using this notation, Eq.~(\ref{eq:greensol}) can be written as
\begin{equation}
G(z,z') = \left\{\begin{array}{ll}-\beta'(z') \alpha(z)& z<z'\\
         -\beta'(z') \alpha(z') & z>z'\end{array}\right. ,
\end{equation}
and $V_1(z)$ as
\begin{equation}
V_1(z) = \int_\epsilon^z \beta(z') \alpha'(z') dz'.
\end{equation}
Having determined the functions $V_n$ in terms of the Green's
function and $V_{n-1}$, let us turn our attention to the
polarization function and its moments. As before,
Eq.~(\ref{eq:polar}) relates the polarization function to the
vector field. Using the expansion in $Q^2$ for $V(z,Q^2)$ in
Eq.~(\ref{eq:series}), we can expand $\Pi^{(c)}$ in powers of
$Q^2$,
\begin{equation} \label{eq:piexp}
\Pi^{(c)}=  \frac{1}{g_5^2}
\sum_{n=-1}\left(-Q^2\right)^{n}\left(e^{B(z)} V'_{n+1}(z)
\right)_{z \rightarrow \epsilon}.
\end{equation}
One may worry since this expansion seems to have a $(Q^2)^{-1}$
term. However, upon further inspection, the coefficient of this
term is $\left.e^{B(z)}V'_0(z)\right|_{z \rightarrow \epsilon}$,
and with the previous assessment that $V_0(z)=1$, this term is
exactly $0$. The expansion also contains a term constant in $Q^2$.
This term can always be eliminated by renormalizing the
polarization function such that $\Pi^{(c)}(0)=0$.

The moments can be calculated in the normal manner from
Eq.~(\ref{eq:moment}) and yield
\begin{equation} \label{eq:momentsol}
\begin{split}
\mathscr{M}_n &= \frac{1}{g_5^2}
\left(e^{B(z)}V'_{n+1}(z)\right)_{z\rightarrow\epsilon}\\
&= \frac{1}{g_5^2} \,  e^{B(\epsilon)}\!\!
\int_\epsilon^\infty\!\! G'(\epsilon,z')
V_n(z') dz'\\
&= \frac{1}{g_5^2}  \int_\epsilon^\infty e^{B(z')} V_n(z') dz',
\end{split}
\end{equation}
where Eqs.~(\ref{eq:sol}) and (\ref{eq:gprime}) are used for lines
two and three respectively. The moment can be expressed using the
compact notation as
\begin{equation}
\mathscr{M}_n  =\frac{1}{g_5^2} \int_\epsilon^\infty\!\!\beta(z)
    V'_n(z) dz.
\end{equation}
The first moment can be explicitly calculated as
\begin{equation}
\begin{split}
 \mathscr{M}_1 &= \frac{1}{g_5^2}
\int_\epsilon^\infty
e^{B(z)}V_1(z) dz\\
    &=\frac{1}{g_5^2} \int_\epsilon^\infty\!\! e^{B(z)} \left[
    \int_\epsilon^z \!\!\left(e^{-B(z')}\!\!\int_{z'}^\infty\!\! e^{B(z'')}
    dz''\right)dz'\right]dz\\
    &=\frac{1}{g_5^2}\int_\epsilon^\infty dz\, e^{-B(z)} \left(\int_z^\infty e^{B(z')}
    dz'\right)^2,
\end{split}
\end{equation}
or in the compact notation,
\begin{equation}
\mathscr{M}_1 = \frac{1}{g_5^2} \int_\epsilon^\infty\!\!
\beta(z)^2 \alpha'(z) dz.
\end{equation}
All higher moments can then be found by iteratively solving
Eq.~(\ref{eq:eom2}) for $V_n(z)$ and using it in
Eq.~(\ref{eq:momentsol}). One can easily see that, for this class
of holographic models, $\mathscr{M}_n$ is positive for all values
of $n$.

\end{appendix}


\begin{thebibliography}{99}

\bibitem{Grigoryan:2010pj}
  H.~R.~Grigoryan, P.~M.~Hohler and M.~A.~Stephanov,
  arXiv:1003.1138 [hep-ph].


\bibitem{Shifman:1978bx}
  M.~A.~Shifman, A.~I.~Vainshtein and V.~I.~Zakharov,
  Nucl.\ Phys.\  B {\bf 147}, 385 (1979).

\bibitem{Shifman:1978by}
  M.~A.~Shifman, A.~I.~Vainshtein and V.~I.~Zakharov,
  Nucl.\ Phys.\  B {\bf 147}, 448 (1979).

\bibitem{Novikov:1977dq}
  V.~A.~Novikov, L.~B.~Okun, M.~A.~Shifman, A.~I.~Vainshtein, M.~B.~Voloshin and V.~I.~Zakharov,
  Phys.\ Rept.\  {\bf 41}, 1 (1978).



\bibitem{Fujita:2009wc}
  M.~Fujita, K.~Fukushima, T.~Misumi and M.~Murata,
  Phys.\ Rev.\  D {\bf 80}, 035001 (2009).

\bibitem{Fujita:2009ca}
  M.~Fujita, K.~Fukushima, T.~Kikuchi, T.~Misumi and M.~Murata,
  arXiv:0911.2298 [hep-ph].

\bibitem{Reinders:1984sr}
  L.~J.~Reinders, H.~Rubinstein and S.~Yazaki,
  Phys.\ Rept.\ {\bf 127}, 1 (1985).




\bibitem{Maldacena:1997re}
  J.~M.~Maldacena,
  Adv.\ Theor.\ Math.\ Phys.\  {\bf 2}, 231 (1998)
  [Int.\ J.\ Theor.\ Phys.\  {\bf 38}, 1113 (1999)];
%

\bibitem{Gubser:1998bc}
  S.~S.~Gubser, I.~R.~Klebanov and A.~M.~Polyakov,
  Phys.\ Lett.\ B {\bf 428}, 105 (1998);

\bibitem{Witten:1998qj}
E.~Witten,
  Adv.\ Theor.\ Math.\ Phys.\  {\bf 2}, 253 (1998).

\bibitem{Erlich:2005qh}
  J.~Erlich, E.~Katz, D.~T.~Son and M.~A.~Stephanov,
  Phys.\ Rev.\ Lett.\  {\bf 95}, 261602 (2005).

\bibitem{Da Rold:2005zs}
  L.~Da Rold and A.~Pomarol,
  Nucl.\ Phys.\  B {\bf 721}, 79 (2005)
  [arXiv:hep-ph/0501218].

\bibitem{Hirn:2005vk}
  J.~Hirn, N.~Rius and V.~Sanz,
  Phys.\ Rev.\  D {\bf 73}, 085005 (2006)
  [arXiv:hep-ph/0512240].

\bibitem{Csaki:2006ji}
  C.~Csaki and M.~Reece,
  JHEP {\bf 0705}, 062 (2007)
  [arXiv:hep-ph/0608266].

\bibitem{Afonin:2009pd}
  S.~S.~Afonin,
  Phys.\ Lett.\  B {\bf 678}, 477 (2009)
  [arXiv:0902.3959 [hep-ph]].

\bibitem{Cappiello:2009cj}
  L.~Cappiello and G.~D'Ambrosio,
  arXiv:0912.3721 [hep-ph].

\bibitem{Hambye:2006av}
  T.~Hambye, B.~Hassanain, J.~March-Russell and M.~Schvellinger,
  Phys.\ Rev.\  D {\bf 76}, 125017 (2007)
  [arXiv:hep-ph/0612010].




\bibitem{Karch:2006pv}
  A.~Karch, E.~Katz, D.~T.~Son and M.~A.~Stephanov,
  Phys.\ Rev.\ D {\bf 74}, 015005 (2006).

\bibitem{Son:2002sd}
  D.~T.~Son and A.~O.~Starinets,
  JHEP {\bf 0209}, 042 (2002).

\bibitem{Herzog:2002pc}
  C.~P.~Herzog and D.~T.~Son,
  JHEP {\bf 0303}, 046 (2003)
  [arXiv:hep-th/0212072].

\bibitem{Son:2003et}
  D.~T.~Son and M.~A.~Stephanov,
  Phys.\ Rev.\  D {\bf 69}, 065020 (2004)
  [arXiv:hep-ph/0304182].




\bibitem{pdg}
 C.~Amsler et al. (Particle Data Group), Physics Letters B667, 1 (2008) and
2009 partial update for the 2010 edition.



\end{thebibliography}
\end{document}